\documentstyle[12pt]{article}
\newcommand{\be}{\begin{equation}}
\newcommand{\ee}{\end{equation}}
\begin{document}

\begin{center}
{\Large\bf Simulations of Quantum Logic Operations \\
in Quantum Computer with Large Number of Qubits}\\ \ \\
G.P. Berman$^1$, G.D. Doolen$^1$, G.V. L\'opez$^2$, and V.I. Tsifrinovich$^3$\\ \end{center}
\ \\
$^1$Theoretical Division and CNLS, 
Los Alamos National Laboratory, Los Alamos,\\
 New Mexico 87545\\
$^2$ Departamento de F\'isica, Universidad de Guadalajara,
Corregidora 500, S.R. 44420, Guadalajara, Jalisco, M\'exico\\
$^3$Department of Physics, Polytechnic University,
Six Metrotech Center, Brooklyn NY 11201\\ \ \\
\begin{center}
{\bf ABSTRACT}
\end{center}
We report the first simulations of the dynamics of quantum logic operations with a large number of qubits (up to 1000). A nuclear spin chain in which selective excitations of spins is provided by the gradient of the external magnetic field is considered. The spins interact with their nearest neighbors. We simulate the quantum CONTROL-NOT (CN) gate implementation for remote qubits which provides the long-distance entanglement. Our approach can be applied to any implementation of quantum logic gates involving a large number of qubits.
\newpage
\quad\\
{\bf 1.~Introduction}\\ \ \\
The field of quantum computation has achieved three important milestones: the first quantum algorithm \cite{1}, 
the first error correction code \cite{2},
and the first experimental implementation of quantum logic \cite{3}. The next promising important step is implementation of quantum logic in solid-state systems
with large number of qubits, say 1000 qubits. It is not clear which system will be the most feasible for quantum computation: nuclear spins \cite{4}-\cite{6}, electron spins \cite{7}-\cite{9}, quantum dots \cite{10}, or Josephson junctions \cite{11}-\cite{12}. For all of these implementations, the design of a quantum computer requires simulations of the quantum computation dynamics on a conventional digital computer to test the quantum computer experimental devices. 

To simulate a general quantum computation involving $N$ qubits, one must solve time-dependent equations involving all $2^N$ states of a quantum computer. Generally, this problem cannot be solved on the digital computer. However, it is possible to simulate quantum logic operations which involve a limited number of states. These simulations can give insight into the dynamical properties of a quantum computer. Simulations of experimental implementations of quantum logic operations can explore the advantages and disadvantages of experimental devices long before they are built. In this paper, we report the first simulation of quantum logic for a large number of qubits (up to 1000). In Sec. 2, we describe the nuclear spin quantum computer which we simulate. In Sec. 3, we consider the Hamiltonian of the nuclear spin chain and the equations of motion for the amplitudes of the quantum states. In Sec. 4, we discuss resonant and non-resonant transitions in the spin chain under the action of radio-frequency ({\it rf}) pulses. In Sec. 5, we describe a quantum CN gate which entangles the two qubits at opposite ends of the spin chain.  In Sec. 6, we give analytical analysis of the CN gate. In Sec. 7, the results of our simulations are presented. In the Conclusion we summarize our results.\\ \ \\
{\bf 2.~Nuclear spin quantum computer}\\ \ \\

We consider a chain of identical nuclear spins placed in a high external magnetic field, $B_0$ (Fig. 1). We suppose that these spins are initially polarized along the direction of the external field ($z$-direction). The NMR frequency is $f_0=(\gamma/2\pi)B_0$, where $\gamma$ is the nuclear gyromagnetic ratio. For example, for a proton in the field $B_0=10$T, one has the NMR frequency $f_0\approx 430$MHz. 

Next, we suppose that the external magnetic field is slightly non-uniform, $B_0=B_0(z)$. Suppose that the frequency difference of two neighboring spins is, $\Delta f\approx 10$kHz. Thus, if the frequency of the edge spin is $430$MHz, the frequency of the other edge spin is $\approx 440$MHz. Then, the value of $B_0$ increases by $\Delta B_0=0.23$T along the spin chain. Taking the distance between the neighboring spins, $a\approx 2\AA$, we obtain the value for the gradient of the magnetic field, $|\partial B_0/\partial z|\approx 0.23/1000 a\cos\theta$, where $\theta$ is the angle between the direction of the chain and the $z$-axis (Fig. 1). Below we will take $\cos\theta=1/\sqrt{3}$. Thus, the gradient of the magnetic field is $|\partial B_0/\partial z|\approx 2\times 10^6$T/m.

Next, we discuss the interaction between spins. In a large external magnetic field, $B_0$, the stationary states of the chain can be described as a combination of individual states of nuclear spins, for example,
$$
|00...00\rangle,~|00...01\rangle, 
$$
and so on, where the state $|0\rangle$ corresponds to the direction of a nuclear spin along the direction of the magnetic field and the state $|1\rangle$ to the opposite direction. The magnetic dipole field on nucleus $j$ in any stationary state is much less than the external field. So, only the $z$-component of the dipole field, $B_{dz}=B_{dz}(j)$, affects the energy spectrum,
$$
B_{dz}(j)=\sum_{k=0}^{N-1}{{3\cos^2\theta-1}\over{r_{kj}^3}}\mu_{kz},~(k\neq j),\eqno(1)
$$
where $\mu_{kz}$ is the $z$-component of the nuclear magnetic moment, $r_{kj}$ is the distance between the nuclei $k$ and $j$. To suppress the dipole interaction, one should choose the angle $\theta\approx 54.7^o$, for which $\cos\theta=1/\sqrt{3}$. Then, for any stationary state, the $z$-component of the dipole field disappears. 

We assume that main interaction between the nuclear spins (when the dipole interaction is suppressed) is an Ising type of interaction mediated by chemical bonds. This situation is observed in liquids where the dipole-dipole interaction is suppressed by the rotational motion of the molecules. Nuclear spins in liquids were used for quantum computations involving a small number of qubits \cite{13,14}. \\ \ \\
{\bf 3.~The Hamiltonian and equations of motion }\\ \ \\

The Hamiltonian for the chain of spins considered can be written in the form \cite{15},
$$
{\cal H}=-\sum_{k=0}^{N-1}\omega_k I^z_k-2J\sum_{k=0}^{N-2}I^z_kI^z_{k+1}+V,\eqno(2)
$$
where $\omega_k$ is the Larmor frequency of the $k$-th spin (neglecting interactions between spins), $\omega_k=\gamma B_0(z_k)$, $J$ is the constant of the Ising interaction, $I^z_k$ is the operator of the $z$-component of spin 1/2, the operator, $V$, describes the interaction with pulses of the {\it rf} field, $z_k$ is the $z$-coordinate of the $k$-th spin, and we set $\hbar=1$.

Below, we assume that the characteristic values of the parameters in the Hamiltonian (2) are,
$$
\omega_k/2\pi=f_0+k\Delta f,~f_0\approx 430 MHz,~\Delta f\approx 10 kHz, ~
J/2\pi\approx 100 Hz,~ N\le 1000.\eqno(3)
$$

The operator $V$ for the $n$-th {\it rf} pulse can be written \cite{15},
$$
V=-{{\Omega^{(n)}}\over{2}}\sum_{k=0}^{N-1}\Bigg[I^-_k\exp(-i\omega^{(n)}t)+I^+_k\exp(i\omega^{(n)}t)\Bigg],\eqno(4)
$$
where $\Omega^{(n)}$ is the Rabi frequency of the $n$-th pulse, $I^{\pm}_k=I^x_k\pm iI^y_k$, and $\omega^{(n)}$ is the frequency of the $n$-th pulse. We choose the value of $\Omega\sim 0.1$J. (We assume that the {\it rf} field is circularly polarized in the $xy$ plane.)

In the interaction representation, the wave function, $\Psi$,  of the spin chain can be written as,
$$
\Psi=\sum_pC_p|p\rangle\exp(-iE_pt),
$$
where $E_p$ is the energy of the state $|p\rangle$. Substituting the expression for the wave function $\Psi$ into the Schr\"odinger equation, we obtain the equation of motion for the amplitude $C_p$,
$$
i\dot C_p=\sum_{m=0}^{2^N-1}C_mV^{(n)}_{pm}\exp[i(E_{p}-E_m)t+ir_{pm}\omega^{(n)}t],\eqno(5)
$$
where $r_{pm}=\mp 1$ for $E_p>E_m$ and $E_p<E_m$, respectively, $V^{(n)}_{pm}=-\Omega^{(n)}/2$ for the states $|p\rangle$ and $|m\rangle$ which are connected by a single-spin transition, and $V^{(n)}_{pm}=0$ for all other states. \\ \ \\
{\bf 4.~Resonant and non-resonant transitions}\\ \ \\
As the number of spins, $N$, increases, the number of states increases exponentially but the number of resonant frequencies in our system is $3N-2$ because only single-spin transitions are allowed by the operator $V$ in (4).
The resonant frequencies of our spin chain are,
$$
\omega_k\pm J,~(k=0,~or~k=N-1),\eqno(6)
$$
$$
\omega_k,~\omega_k\pm 2J, (1\le k\le N-2).
$$
For edge spins with $k=0$ and $k=N-1$, the upper and lower signs correspond to the states $|0\rangle$ or $|1\rangle$ of the only neighboring spin. For inner spins with $1\le k\le N-2$ the frequency $\omega_k$ corresponds to having nearest neighbors whose spins are in opposite directions to each other.  The ``+'' sign corresponds to 
having the nearest neighbors in their ground state. The ``-'' sign corresponds to 
having the nearest neighbors in their excited  state. 

Now we consider any basic stationary state,
$$
|q_{N-1}q_{N-2}...q_1q_0\rangle,\eqno(7)
$$
where the subscript indicates the position of the spin in a chain, and $q_k=0,1$. If one applies to the spin chain a resonant {\it rf} pulse of a frequency, $\omega$, from (6) one has two possibilities:\\
1). The frequency of the pulse $\omega$ is the resonant frequency of the state (7).\\
2). The frequency $\omega$ differs from the closest resonant frequency of the state (7) by the value $2J$ or $4J$.

In the first case, one has a resonant transition. For the second case, one has a non-resonant transition. If $J\ll 2\pi\Delta f$ we can neglect all other non-resonant transitions for the state (7). Below, we will write a rigorous condition for $\Delta f$ which is required in order to neglect all other non-resonant transitions.

Thus, considering the transformation of any basic state under the action of an {\it rf} pulse with a frequency, $\omega$, from (6), we should take into consideration only one transition. This transition will be either a resonant one or a non-resonant one with the frequency difference $2J$ or $4J$. 

This allows us to simplify equations (5) to the set of two coupled equations,
$$
i\dot C_p=-(\Omega^{(n)}/2)\exp[i(E_p-E_m-\omega^{(n)})t]C_m,\eqno(8)
$$
$$
i\dot C_m=-(\Omega^{(n)}/2)\exp[i(E_m-E_p+\omega^{(n)})t]C_p,
$$
where $E_p>E_m$, $|p\rangle$ and $|m\rangle$ are any two stationary states which are connected by a single-spin transition and whose energies differ by $\omega^{(n)}$ or $\omega^{(n)}\pm 2J$ or  $\omega^{(n)}\pm 4J$.

The solution of equations (8) for the case when the system is initially in a stationary state $|m\rangle$, can be written,
$$
C_m(t_0+\tau)=[\cos(\Omega_e\tau/2)+i(\Delta/\Omega_e)\sin(\Omega_e\tau/2)]\times\exp(-i\tau\Delta/2),\eqno(9)
$$
$$
C_p(t_0+\tau)=i(\Omega/\Omega_e)\sin(\Omega_e\tau/2)\times\exp(it_0\Delta+i\tau\Delta/2),
$$
$$
C_m(t_0)=1,~C_p(t_0)=0.
$$
In Eqs (9) we omitted the upper index ``$n$'' which indicates the number of the {\it rf} pulse, $t_0$ is the time of the beginning of the pulse, $\tau$ is its duration, $\Delta =E_p-E_m-\omega$, $\Omega_e=(\Omega^2+\Delta^2)^{1/2}$ is the NMR frequency in the rotating frame. If the system is initially in the upper state, $|p\rangle$, ($C_m(t_0)=0$, $C_p(t_0)=1$), one can obtain the solution of Eqs (8) by changing the sign at $\Delta$ and setting: $m\rightarrow p$ and $p\rightarrow m$ in (9).

For the resonant transition ($\Delta=0$) the expressions (9) transform  into the well-known equations for the Rabi transitions,
$$
C_m(t_0+\tau)=\cos(\Omega\tau/2),~C_p(t_0+\tau)=i\sin(\Omega\tau/2).\eqno(10)
$$
In particular, for $\Omega\tau=\pi$ (a $\pi$-pulse),  Eqs (10) describe the complete transition from the state $|m\rangle$ to the state $|p\rangle$.

For non-resonant transitions, expressions (9) include two characteristic parameters: $\Omega/\Omega_e$ and $\sin(\Omega_e\tau/2)$. If either of these two parameters is zero, the probability of a non-resonant transition disappears.  The second parameter is equal zero when $\Omega_e\tau=2\pi k$ ($k=1,2,..$), where $k$ is the number of revolutions of a non-resonant (average) spin about the effective field in the rotating frame. 
This is the basis of the ``$2\pi k$''-method for elimination non-resonant transitions. (See \cite{15}, Chapter 22, and \cite{16}.)\\ \ \\
{\bf 5.~ A Control-Not gate involving remote qubits and their long-distance entanglement}\\ \ \\
A pure Control-Not ($CN_{ab}$) gate is a unitary operator which transforms the basic state, 
$$
|q_{N-1}...q_a.......q_b....q_0\rangle,
$$
into the state,
$$
|q_{N-1}...q_a.......\bar q_b....q_0\rangle,
$$
where $\bar q_b=1-q_b$ if $q_a=1$; and $\bar q_b=q_b$ if $q_a=0$. The $a$-th and $b$-th qubits are called the control and the target qubits of the $CN_{ab}$ gate. A modified CN gate performs the same transformation  accompanied by phase shifts which are different for different basic states \cite{15}. It is well-known that the CN gate can produce an entangled state of two qubits, which can not be represented as a product of the individual wave functions.

We shall consider an implementation of the CN gate in the Ising spin chain with the left end spin as the control qubit and the right end spin as the target qubit,
i.e. $CN_{N-1,0}$ for a spin chain of 200 and a spin chain of 1000 qubits. Using this gate we will create entanglement between the end qubits in the spin chain. We start with the ground state. Then we apply a $\pi/2$-pulse with  frequency 
$\omega_{N-1}$ to produce a superpositional state of the $(N-1)$-th (left) qubit,
$$
\Psi=|0...0\rangle+i|1...0\rangle.\eqno(11)
$$
(Here and below the normalization factor $1/\sqrt{2}$ is omitted.) To implement a modified $CN_{N-1,0}$ gate we apply to the spin chain $L=397$ $\pi$-pulses if $N=200$, and $L=1997$ pulses if $N=1000$. The first $\pi$-pulse has the frequency $\omega=\omega_{N-2}$. For the second $\pi$-pulse $\omega=\omega_{N-3}$. For the third $\pi$-pulse, $\omega=\omega_{N-2}-2J$, etc.\\ \ \\
{\bf 6.~ An analytic solution}\\ \ \\
An analytical expression for the wave function, $\Psi$, after the action of $\pi/2$- and $L$ $\pi$-pulses, can be easily derived if for all $\pi$-pulses $\Omega_e\tau/2=2\pi k$, with the same value $k$:
$$
\Psi=C_0|00..0\rangle+C_1|10...1\rangle,\eqno(12)
$$
$$
C_0=(-1)^{kL}\exp(-i\pi L\sqrt{4k^2-1}/2),~C_1=-1,\eqno(13)
$$
 For $k\gg 1$, we get the same solution for odd and even $k$: $C_0\approx 1$. This result is easy to understand. For a $\pi$-pulse, the Rabi frequency is $\Omega=|\Delta|/\sqrt{4k^2-1}$. Large values of $k$ correspond to small values of the parameter $\Omega/|\Delta|$.  If $\Omega/|\Delta|$ approaches zero, the non-resonant pulse cannot change the quantum state.

For a small value of $k$, the non-resonant pulse can change the phase of a state. For example, for $k=1$ we have,
$$
C_0=\exp[i\pi L(1-\sqrt{3}/2)].
$$
After the first $\pi$-pulse, the phase shift is approximately $24^o$, but it grows as the number of $\pi$-pulses, $L$, increases.

Now, we shall mention an important point. If we consider the probability of non-resonant transition, the small parameter of the problem is:
$$
\epsilon=(\Omega/\Omega_e)^2\sin^2(\Omega_e\tau/2).
$$
(It follows from (9) that the expression for $|C_m|^2$ can be written in the form: $|C_m|^2=1-\epsilon$, and $|C_p|^2=\epsilon$.) If we take into consideration the change of the phase of non-resonant state, the small parameter of the problem is $\Omega/\|\Delta|\approx \Omega/\Omega_e$.

Next, we will discuss the probability of non-resonant transitions using perturbation theory. The analytical expression for probabilities $|C_0|^2$ and $|C_1|^2$ can be easily found in the first non-vanishing approximation of perturbation theory:
$$
|C_0|^2=1-L\epsilon,~|C_1|^2=1.\eqno(14)
$$
The decrease of the probability $|C_0|^2$ is caused by the generation of unwanted states.
One can see that the deviation from the value $|C_0|^2=1$ accumulates when the number of $\pi$-pulses, $L$, increases. 
It means that the small parameter of the problem is $L\epsilon$ rather than $\epsilon$. Consider first those non-resonant transitions which we ignore in this paper. For the number of qubits, $N=1000$,  using the characteristic parameters from (3) and $L\approx 2000$, $\Omega\approx 0.1J$, $|\Delta|=2\pi\Delta f$, we obtain, $L\epsilon\le 10^{-3}$. 
Thus, as was already pointed out, one can neglect non-resonant transitions whose frequency differences are of the order of $\Delta f$.

Now we consider non-resonant transitions which are included in our simulations. Setting $|\Delta|=2J$, we obtain: $L\epsilon\le2.5$. Thus, the deviation from the $2\pi k$ condition can produce large distortions from the desired wave function. To study these distortions we have to use computer simulations.\\ \ \\
{\bf 7.~ Computer simulations}\\ \ \\
We have developed a numerical code which allowed us to study the dynamics of quantum states with the probabilities no less than $10^{-6}$ for a spin chain with up to 1000 qubits. (Because we omitted in (11) the normalization factor $1/\sqrt{2}$ all probabilities in this section including in Figs 2-14 are doubled.) The sum of the probabilities of all these states was equal $2-O(10^{-6})$ (the normalization condition).

Next, in Figs 2-14, we present the results of computer simulations with $N=200$ and with $N=1000$ qubits. Fig. 2, shows the probability of the excited unwanted states after implementation of the $CN_{199,0}$ gate, for $N=200$ and $\Omega=0.14$. On the horizontal axis the unwanted states are shown in the order of their generation. A total of $7385$ unwanted states were generated which had the probability, $P\ge 10^{-6}$. (In all Figs 2-14 only the states with $P\ge 10^{-6}$ are taken into account.) The probability distribution of unwanted states clearly contains two ``bands''. One group of these states has the probability, $P\sim 10^{-6}$ (the bold ``line'' near the horizontal axis).  The second group of states has the probability $P\sim 10^{-3}$ (the upper ``curve'' in Fig. 2). Fig. 3, shows an enlargement of the upper ``band'' of the Fig. 2. One can see some sub-structure of this ``band''. Fig. 4, shows the sub-structure in the lower ``band'' of Fig. 2. Figs 5 (a,b) show the sub-structure of the upper and lower ``bands'' shown in Fig. 4. One can see some hierarchy in the structure of the distribution function of generated unwanted states. Figs 6(b-j) show the typical structure of unwanted states of the spin chain. Fig. 6a, shows the ground state of the spin chain (all qubits are in their ground state). The value of  $P$ in Fig. 6, indicates the probability of the states. All states in Figs 6(b-j) belong to the upper ``band'' shown in Fig. 2, i.e. they have a probability, $P\sim 10^{-3}$. It is interesting to note that the group of unwanted states with high probabilities contains the high energy states of the spin chain (many-spin excitations). Typical unwanted states of the lower ``band'' in Fig. 2 (with $P\sim 10^{-6}$) are shown in Figs 7(a-j) and in Figs 8(a-j). It is important to note that typical unwanted states for both groups (Figs 6-8) contain highly correlated spin excitations. Fig. 9 shows the total number of unwanted states (with probability $P\ge 10^{-6}$) and the probability of the ground state, $|C_0|^2$, as a function of the Rabi frequency, $\Omega$. The maximum value of $|C_0|^2$ and the minimal total number of unwanted states correspond to the values of $\Omega$ which satisfy the $2\pi k$-condition for $396$ of the total number of $\pi$-pulses, $397$. (The third $\pi$-pulse does not satisfy a $2\pi k$-condition.)

Next, we have studied the generation of unwanted states for the case when only a group of pulses had values of Rabi frequency which deviated from the $2\pi k$-condition. We changed the value of $\Omega$ for all $\pi$-pulses from $k_1=10$ to $k_2=(10+\Delta k)$. Fig. 10, shows the number of unwanted states and the probability of the ground state, $|C_0|^2$, as a function of the number of distorted pulses, $\Delta k$,  for $N=1000$. The value $\Omega\approx 0.100$ in Fig. 10 corresponds to the $2\pi k$-condition for all pulses (except the $3$-rd $\pi$-pulse), for distorted $\Delta k$ pulses, $\Omega=0.101$. Figs 11 and 12 demonstrate the same quantities (the number of unwanted generated states and $|C_0|^2$) for the case when $\Omega$  is a random parameter for a group of pulses. (Both figures show typical realizations for definite distributions of $\Omega$.) Fig. 13, shows the dependence of the number of unwanted states and $|C_0|^2$ on the location of the group of distorted pulses. Fig. 14, demonstrates the ground state and the examples of unwanted states generated due to the distortion of this group of pulses. One can see again that  high-energy states of the spin chain, with many-spin excitations, are  generated.\\ \ \\
{\bf Conclusion}\\ \ \\
In this paper we presented the results of simulations of quantum Control-Not gate, $CN_{N-1,0}$, between remote qubits, $(N-1)$-st and $0$-th, and the creation of long-distance entanglement in nuclear spin quantum computer having a large number of qubits (up to 1000). A considered quantum computer is a one-dimensional nuclear spin chain placed in a slightly non-uniform magnetic field, and oriented in such direction that the dipole interaction between spins is suppressed. So, the Ising interaction comes into play, as in the case of the liquid NMR. We used two essential assumptions:\\
1. The nuclear spin chain is prepared initially in the ground state. \\
2. The frequency difference between the neighboring spins due to the inhomogeneity of the external magnetic field is much large than the Ising interaction constant.

Using these assumptions, we developed a numerical method which allowed us to simulate the dynamics of quantum logic operations taking into consideration all quantum states with the probability no less that $P=10^{-6}$. For the case, when the $2\pi k$-condition is satisfied (the $\pi$-pulse for the resonant transition is at the same time a $2\pi k$-pulse for non-resonant transitions), we obtained an analytic solution for the evolution of the nuclear spin chain. In the case of small deviations from the $2\pi k$-condition, the error accumulates. So, the perturbation theory becomes invalid even for small deviations from the $2\pi k$-condition. In this case, the numerical simulations are necessary.

The main results of our simulations are the following:\\
1. The unwanted states exhibit band structure in their probability distributions. There are two main ``bands'' in the probability distribution of unwanted states. The unwanted states in these ``bands'' have significantly different probabilities, $P_{low}/P_{upper}\sim 10^{-3}$. Each of these two bands have their own structure. \\
2. A typical unwanted state is a state of highly correlated spin excitations. An important fact is that the unwanted states with relatively {\it high
probability} include high energy states of the spin chain (many-spin excitations).\\
3. The method developed allowed us to study generation of unwanted states and the probability of the desired states as a function of the distortion of {\it rf} pulses. This can be used to formulate the requirements for  acceptable errors in quantum computation.

The results of this paper can be used to design experimental implementations of quantum logic operations and to estimate (benchmark) the quality of experimental quantum computer devices. Our approach can be extended to any types of quantum computer.\\ \ \\  
{\bf Acknowledgments}\\ \ \\
This work  was supported by the Department of Energy under contract W-7405-ENG-36 and by the National Security Agency.
\newpage
\quad\\
{\bf Figure captions}\\ \ \\
Fig. 1:~Nuclear spin quantum computer (the ground state of nuclear spins). $B_0$ is the permanent magnetic field; $B_1$ is the radio-frequency field. The chain of spins makes the angle $\theta$ with the direction of the field $B_0$.\\ \ \\
Fig. 2:~Probabilities of unwanted states. The total number of qubits: $N=200$;  
$\Omega=0.14$. The number of unwanted states with probabilities $|C_n|^2\ge 10^{-6}$ 
is $7385$. The states are presented in the order of their generation. \\ \ \\
Fig. 3:~The upper band of Fig. 2, shown in a larger scale.\\ \ \\
Fig. 4:~The lower band of Fig. 2, shown in a larger scale.\\ \ \\
Fig. 5a:~A sub-structure of the upper band of Fig. 4, shown in a larger scale.\\ \ \\
Fig. 5b:~A sub-structure of the lower band of Fig. 4, shown in a larger scale.\\ \ \\
Fig. 6:~ (a) The ground state of the spin chain; (b-j) The typical unwanted states with probabilities $\sim 10^{-3}$. Horizontal axis shows the position of a qubit in the spin chain of $N=200$ spins. The vertical axis shows the state $|0\rangle$ or $|1\rangle$ of the qubit.\\ \ \\
Fig. 7:~ Examples of ``low energy'' unwanted states from the lower band in Fig. 2.\\ \ \\
Fig. 8:~ Examples of ``intermediate energy'' unwanted states from the lower band in Fig. 2.\\ \ \\
Fig. 9:~ (a) Probability, $|C_0|^2$, as a function
of $\Omega$. The total number of qubits $N=200$; (b) The total number of unwanted states.\\ \ \\
Fig. 10:~Dependence of the total number of unwanted states as a  function of $\Delta k=k_2-k_1$ ($k_1=10$, $N=1000$); (a) The total number of unwanted states; $\Omega=0.1$ for all $\pi$-pulses except for the $\pi$-pulses with numbers $k$ in the range: $k_1<k<k_2$ for which $\Omega=0.101$; (b) The probability, $|C_0|^2$,  for the same parameters as in (a).\\ \ \\
Fig. 11:~ (a) The probability, $|C_0|^2$, as a function of parameter $\varepsilon_0$; $N=1000$; (b) The number of unwanted states as a function of parameter $\varepsilon_0$; $\Omega=0.1$ for all pulses from 10-th to $(10+\Delta k)$-th
for which $\Omega=0.1+\varepsilon$; $\varepsilon$ is a random parameter: $\varepsilon\in[-\varepsilon_0,\varepsilon_0]$.\\ \ \\
Fig. 12:~ (a) The probability, $|C_0|^2$, as a  function of $\Delta k=k_2-k_1$; (b) The total number of unwanted states; $\Omega=0.1$ for all $\pi$-pulses except for the $\pi$-pulses from 10-th to $(10+\Delta k)$-th for which $\Omega=0.1+\varepsilon$; $\varepsilon$ is a random parameter in the range $-0.05<\varepsilon<0.05$, $N=1000$. \\ \ \\
Fig. 13:~ (a) The number of unwanted states as a function of the parameter $k_1$; $\Omega=0.1$ for all $\pi$-pulses except for the $\pi$-pulses from $k_1$-th to $(k_1+15)$-th for which $\Omega=0.1+\varepsilon$; $\varepsilon$ is a random parameter, $\varepsilon\in [-0.005,0.005]$; (b) The probability, $|C_0|^2$, for the same parameters as in (a); $N=1000$.\\ \ \\
Fig. 14:~(a) The ground state of the chain; (b-j) Examples of unwanted states ($N=1000$); $\Omega=0.1$ for all $\pi$-pulses 
except for the $\pi$-pulses from 10-th to 40-th 
for which $\Omega=0.1+\varepsilon$, where $\varepsilon$ is a random parameter in the range: 
$-0.05<\varepsilon< 0.05$ .\\ \ \\
\newpage

\begin{thebibliography}{99}
%
\bibitem{1}
P. Shor, {\it Proc. of the 35th Annual Symposium on the Foundations of Computer Science}, {\it IEEE, Computer Society Press}, New York, (1994), p. 124.
%
\bibitem{2}
P. Shor, {\it Phys. Rev. A}, {\bf 1995} R2493.
%
\bibitem{3}
C. Monroe, D.M. Meekhov, B.E. King, W.M. Itano, D.J. Wineland, {\it Phys. Rev. Lett.}, {\bf 75} (1995) 4714. 
%
\bibitem{4}
F. Yamaguchi, Y. Yamamoto, {\it Microelectronic Engineering}, {\bf 47} (1999) 273.
%
\bibitem{5}
B.E. Kane, {\it Nature}, {\bf 393}, (1998) 133.
%
\bibitem{6}
G.P. Berman, G.D. Doolen C.P. Hammel, V.I. Tsifrinovich, cond-mat, 1999.
%
\bibitem{7}
G.P. Berman, G.D. Doolen, G.D. Holm, V.I. Tsifrinovich, {\it Phys. Lett. A}, {\bf 193} (1994) 444.
%
\bibitem{8}
R. Vrijen, E. Yablonovich, K. Wang, H.W. Jiang, A. Balandin, V. Roychwdhury, T. Mor, D. DiVincenzo, quant-ph/9905096 v2.
%
\bibitem{9}
A. Imamoglu, D.D. Awschalom, G. Burkard, D.P.  DiVincenzo, D. Loss, M. Sherwin, A. Small, quant-ph/9904096 v2.
%
\bibitem{10}
M.S. Sherwin, A. Imamoglu, T. Montroy,  quant-ph/9903065.
%
\bibitem{11}
A. Shnirman, G. Schoen, Z. Hermon, {\it Phys. Rev. Lett.}, {\bf 79} (1997) 2371.
%
\bibitem{12}
D.V. Averin, {\it Solid State Coom.}, {\bf 105}, (1998) 659.
%
\bibitem{13}
N.A. Gershenfeld, I.L. Chuang, {\it Science}, {\bf 275}, (1997) 350.
%
\bibitem{14}
D.G. Cory, A.F. Fahmy, T.F. Havel, {\it Proc. Natl. Acad. Sci. USA}, {\bf 94}, (1997) 1634.
%
\bibitem{15}
G.P. Berman, G.D. Doolen, R. Mainieri, V.I. Tsifrinovich, {\it Introduction to Quantum Computers}, World Scientific Publishing Company, 1998.
%
\bibitem{16}
G.P. Berman, D.K. Campbell, V.I. Tsifrinovich, {\it Phys. Rev. B},  {\bf 55}, (1997) 5929.
%
\end{thebibliography}
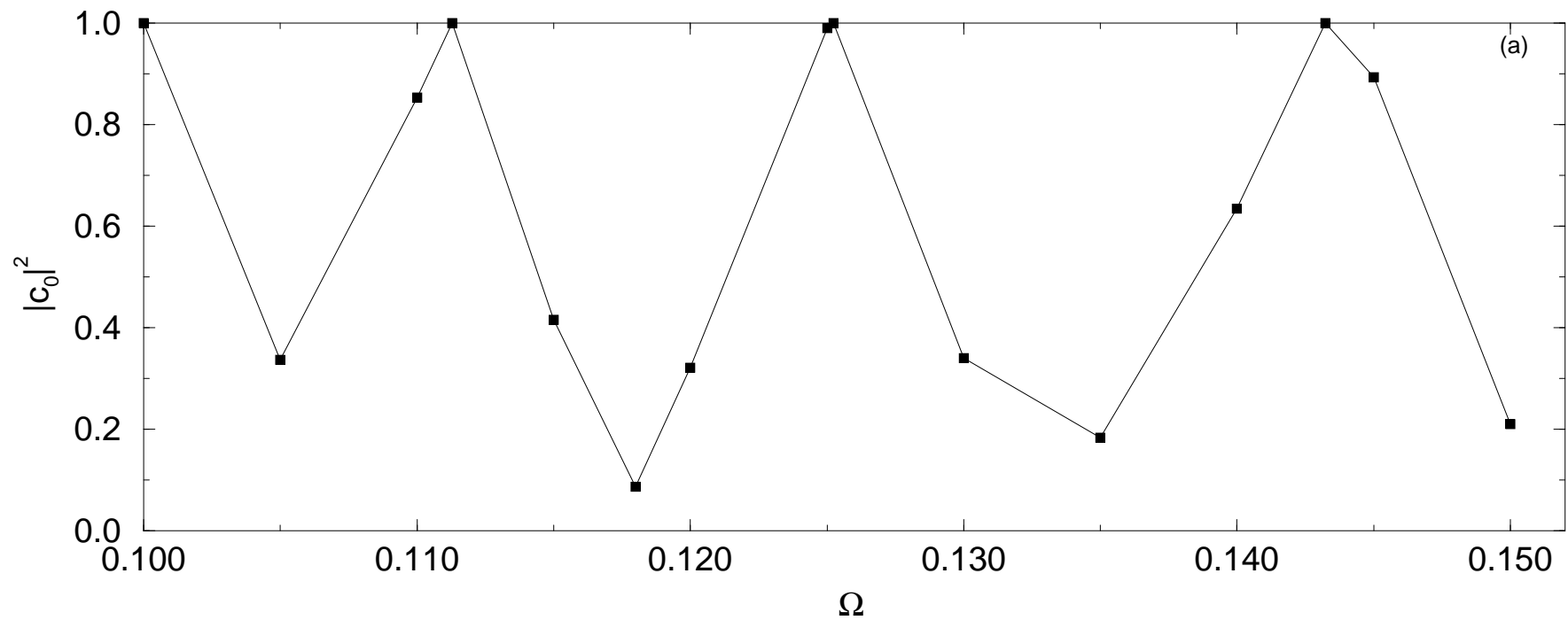
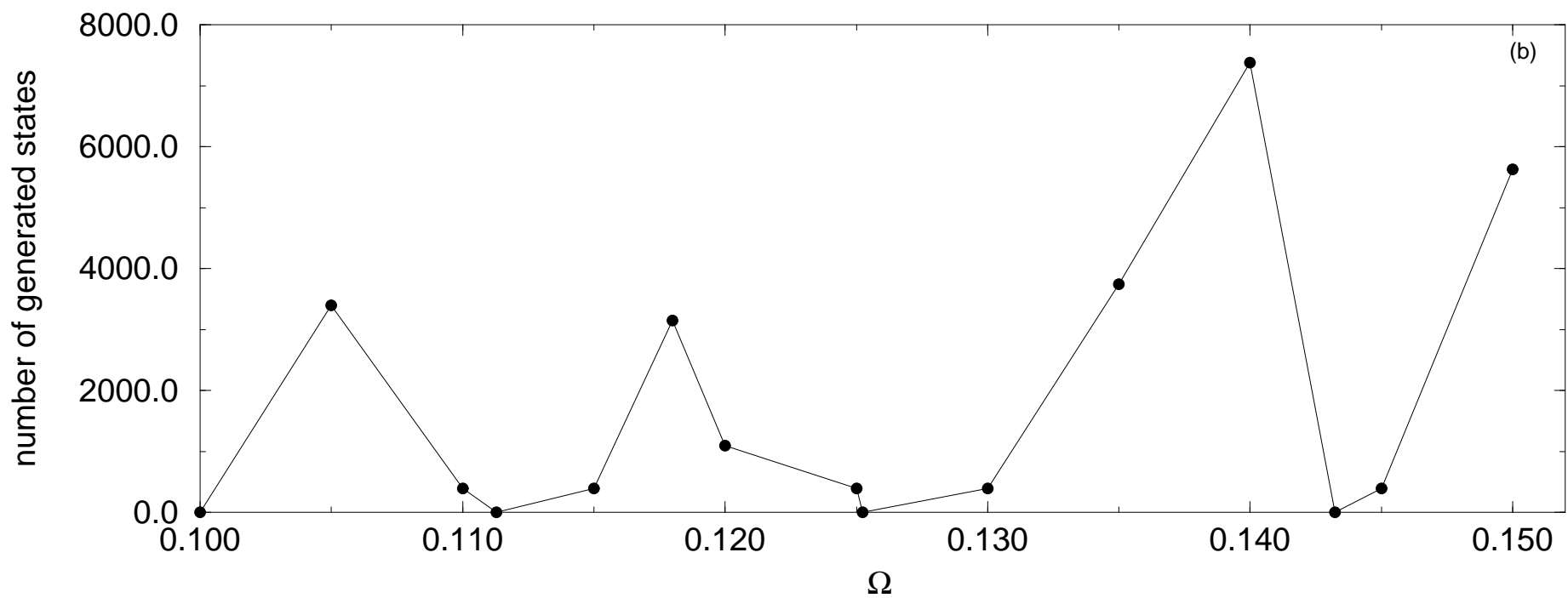
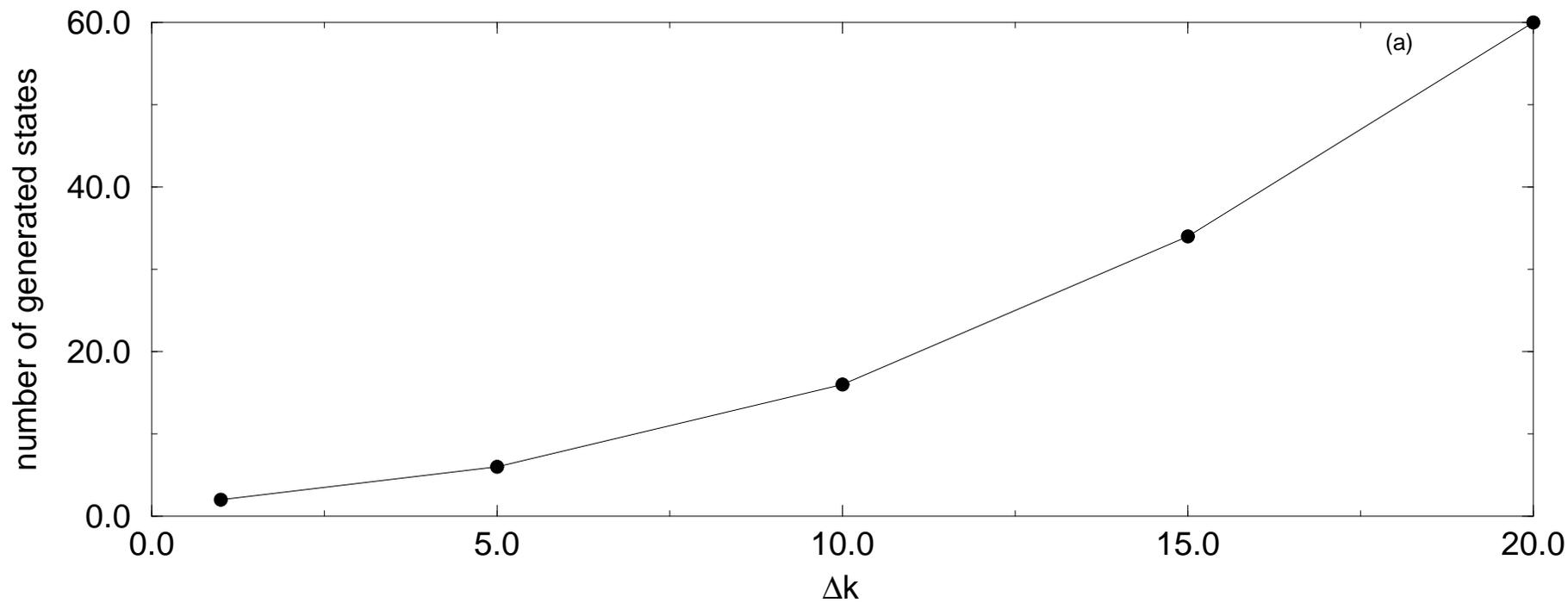
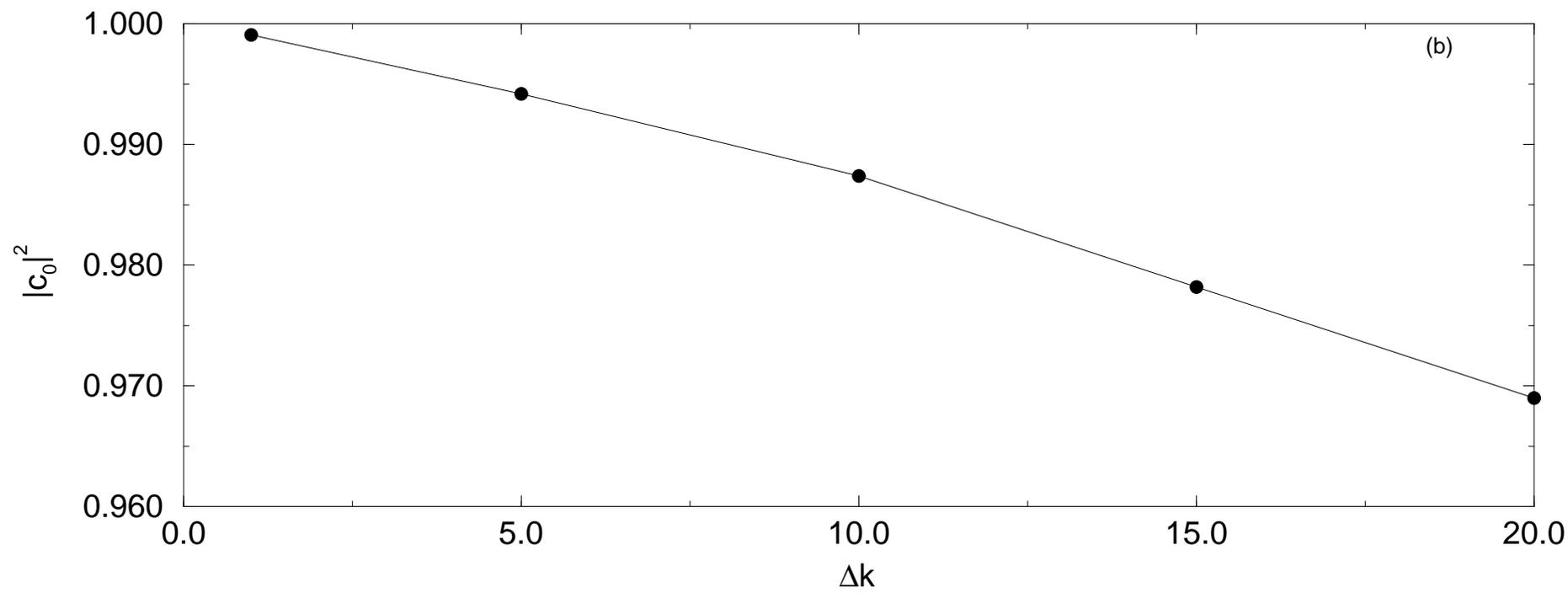
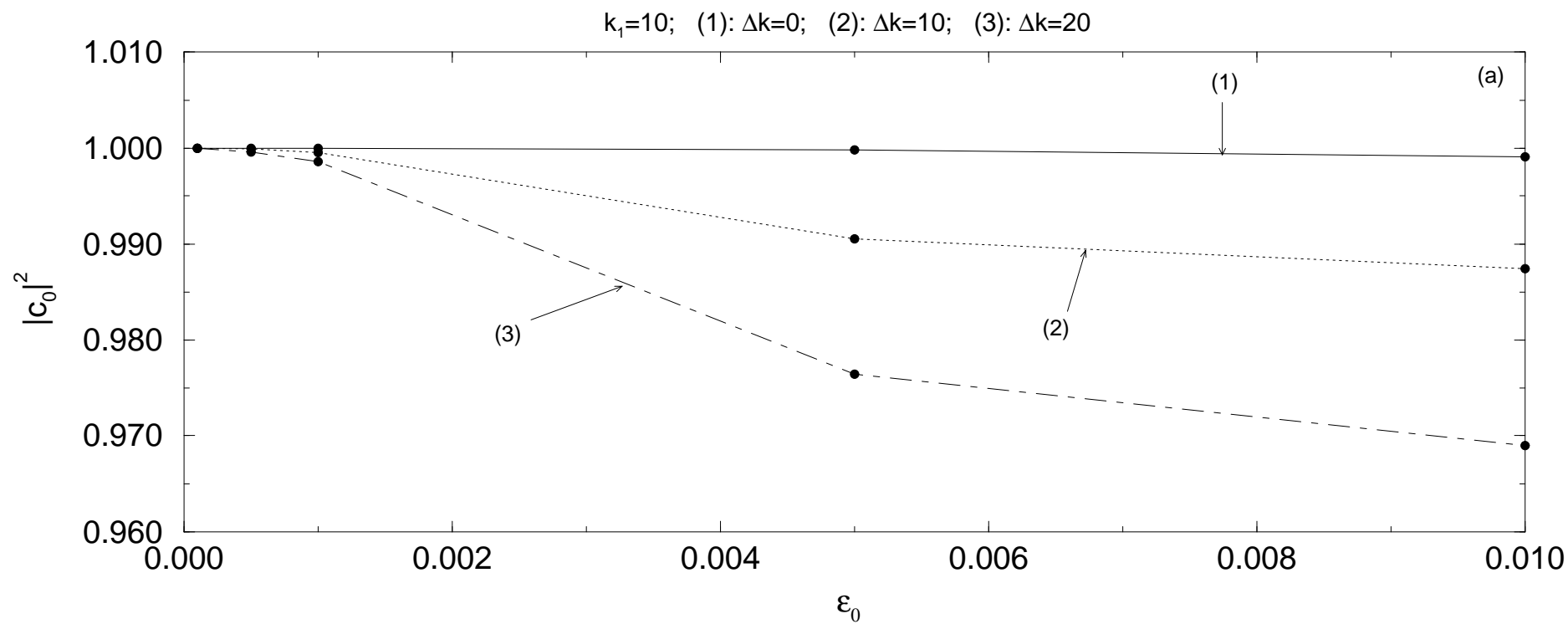
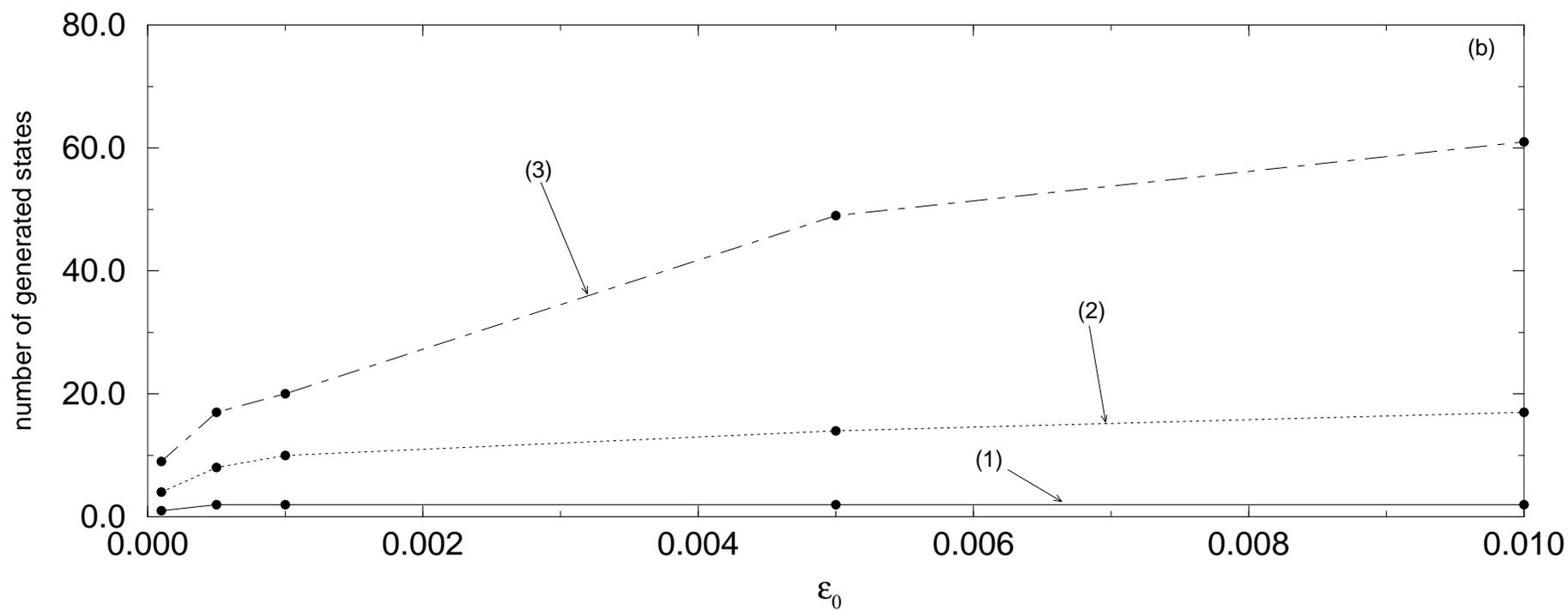
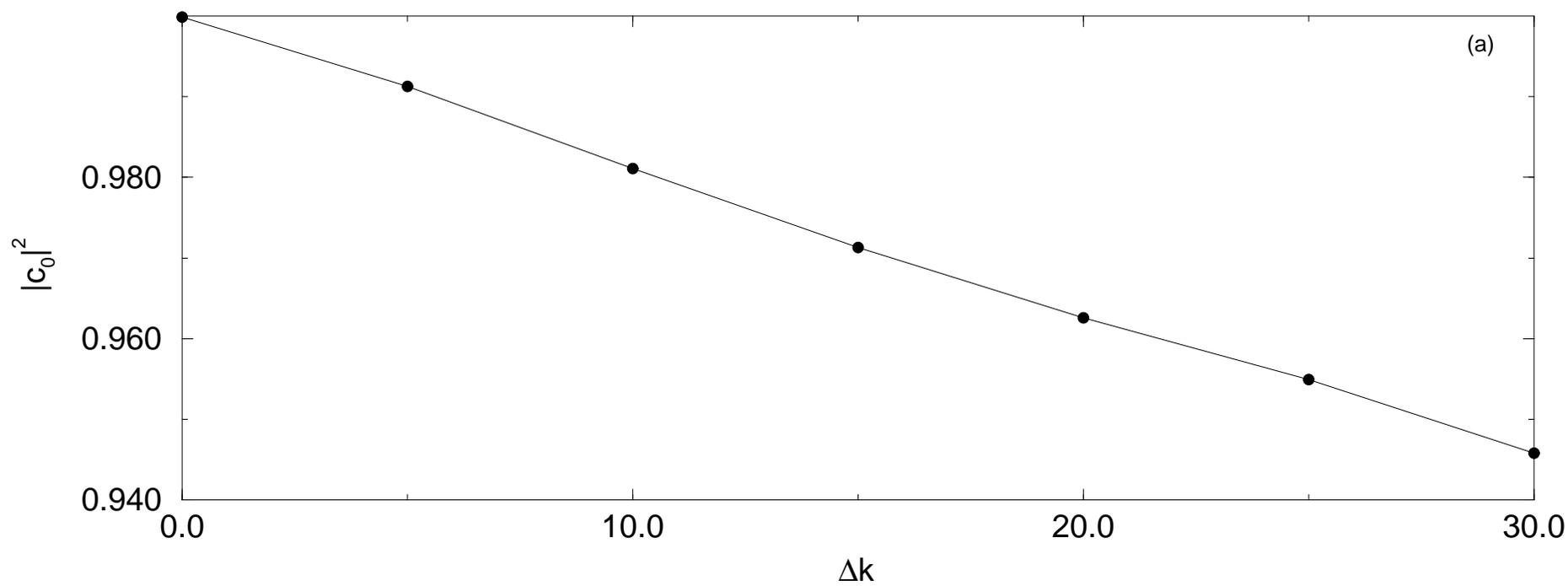
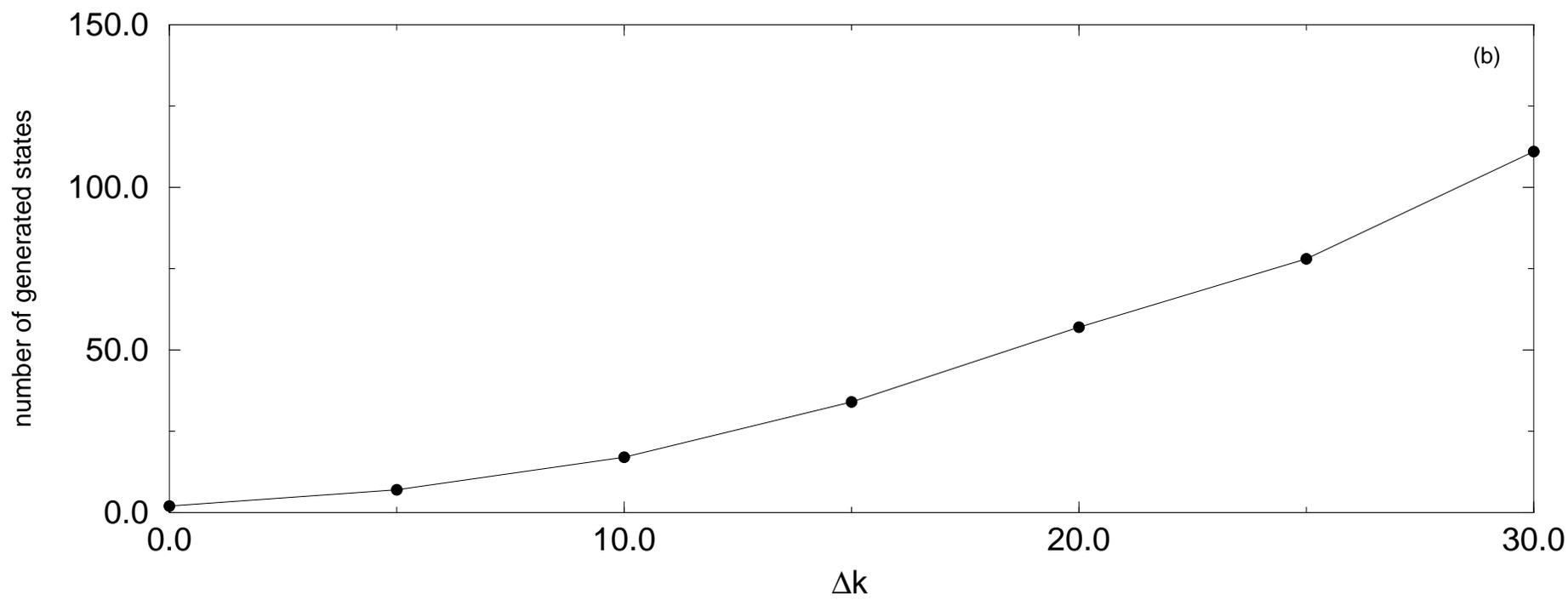
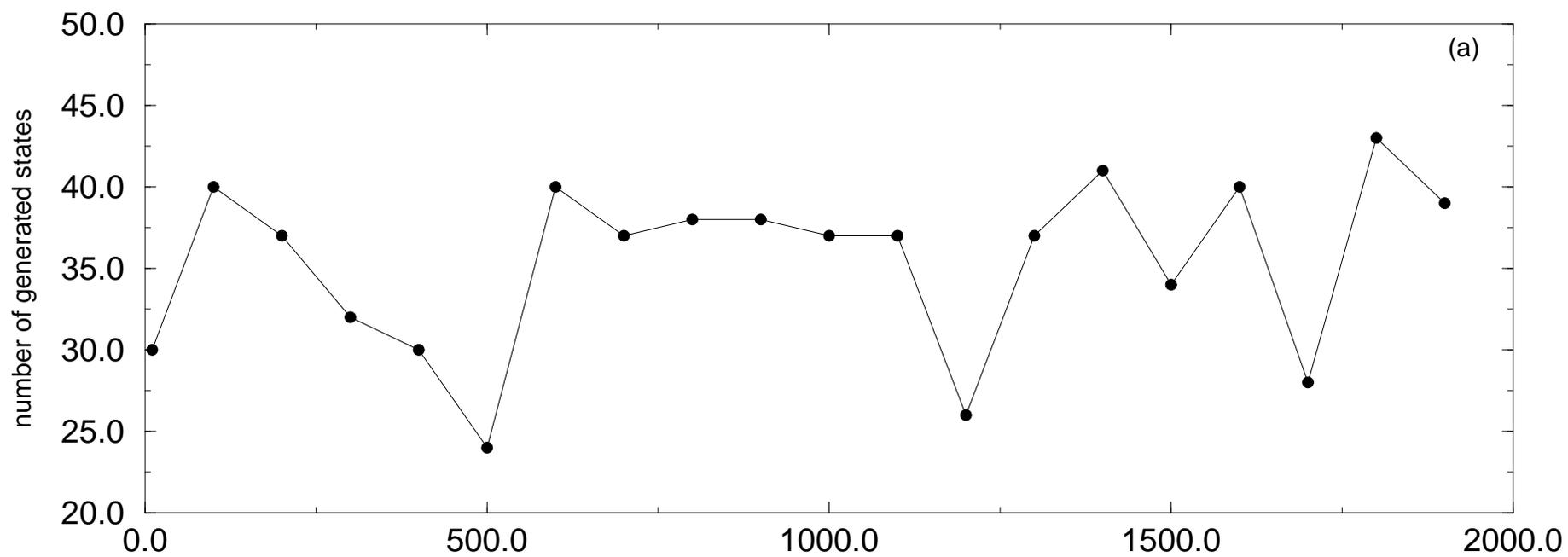
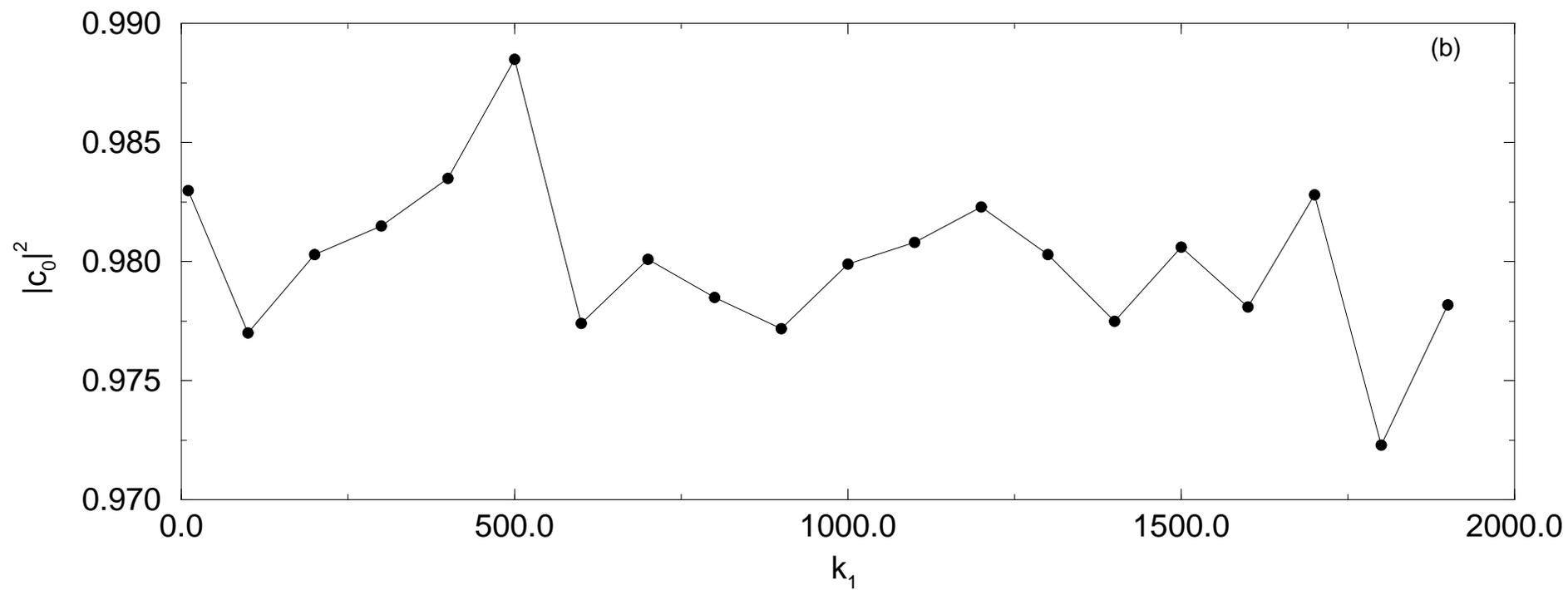
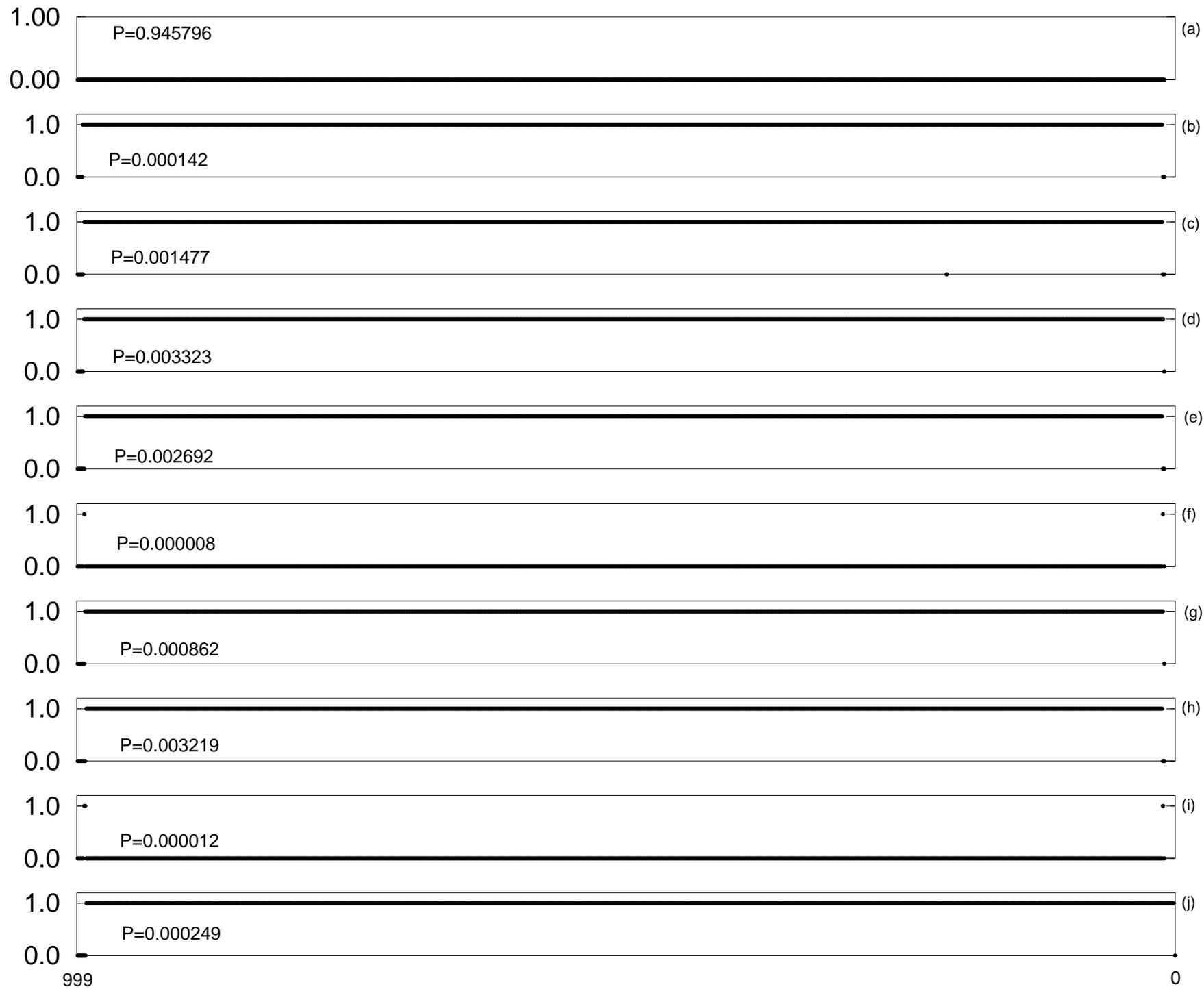
\end{document}